\documentclass[reprint, superscriptaddress, amsmath,amssymb, aps, prl, floatfix]{revtex4-1}

\usepackage[utf8]{inputenc}
\usepackage{graphicx}
\usepackage{dcolumn}
\usepackage{bm}
\usepackage{hyperref}
\usepackage{braket}
\usepackage{booktabs,tabularx}
\newcolumntype{Y}{>{\centering\arraybackslash}X}
\newcolumntype{Z}{>{\raggedleft\arraybackslash}X}

\usepackage{todonotes,xcolor}
\usepackage[normalem]{ulem}

\renewcommand{\d}{{\rm d}}

\newcommand{\sigmaA}{\sigma^\mathrm{A}}

\newcommand{\sigmaAt}{\sigma^\mathrm{A} (t)}
\newcommand{\sigmaAtot}{\sigma^\mathrm{A}_{\rm 300K}}

\newcommand{\kaigk}{\kappa_\mathrm{aiGK}}

\newcommand{\Nexp}{{24}}
\newcommand{\Nnew}{{34}}
\newcommand{\Ntotal}{58}

\newcommand{\Nscreening}{465}

\newcommand{\Nten}{28}

\newcommand{\None}{6}

\newcommand{\Nultra}{\None}

\usepackage{glossaries}
\setacronymstyle{long-short}
\newacronym{gk}{GK}{Green-Kubo}
\newacronym{aigk}{aiGK}{\emph{ab initio} Green-Kubo}
\newacronym{md}{MD}{molecular dynamics}
\newacronym{aimd}{aiMD}{\emph{ab initio} molecular dynamics}
\newacronym{pes}{PES}{potential-energy surface}
\newacronym{bte}{BTE}{Boltzmann transport equation}
\newacronym{sm}{SM}{supplemental material}
\newacronym{hfacf}{HFACF}{heat flux autocorrelation function}
\newacronym{cui}{CuI}{copper iodide}
\newacronym{snse}{SnSe}{tin selenide}
\newacronym{agals2}{AgAlS$_2$}{silver aluminum disulfilde}
\newacronym{mape}{MAPE}{mean absolute percentage error}
\newacronym{mae}{MAE}{mean absolute error}

\newcommand{\agals}{AgAlS$_2$}
\newcommand{\agalse}{AgAlSe$_2$}
\newcommand{\liinte}{LiInTe$_2$}
\newcommand{\ligate}{LiGaTe$_2$}
\newcommand{\kcaf}{KCaF$_3$}

\begin{document}

\title{Anharmonicity in Thermal Insulators -- An Analysis from First Principles}

\author{Florian Knoop}
\affiliation{The NOMAD Laboratory at the FHI of the Max-Planck-Gesellschaft and IRIS-Adlershof of the Humboldt-Universität zu Berlin, Faradayweg 4-6, 14195 Berlin, Germany}
\affiliation{Theoretical Physics Division, Department of Physics, Chemistry and Biology (IFM), Linköping University, SE-581 83 Linköping, Sweden}

\author{Thomas A.\,R. Purcell}
\affiliation{The NOMAD Laboratory at the FHI of the Max-Planck-Gesellschaft and IRIS-Adlershof of the Humboldt-Universität zu Berlin, Faradayweg 4-6, 14195 Berlin, Germany}

\author{Matthias Scheffler}
\affiliation{The NOMAD Laboratory at the FHI of the Max-Planck-Gesellschaft and IRIS-Adlershof of the Humboldt-Universität zu Berlin, Faradayweg 4-6, 14195 Berlin, Germany}

\author{Christian Carbogno}
\affiliation{The NOMAD Laboratory at the FHI of the Max-Planck-Gesellschaft and IRIS-Adlershof of the Humboldt-Universität zu Berlin, Faradayweg 4-6, 14195 Berlin, Germany}

\begin{abstract}
The anharmonicity of atomic motion limits the thermal conductivity in crystalline solids. However, a microscopic understanding of the mechanisms 
active in strong thermal insulators is lacking.
In this letter, we classify \Nscreening{} experimentally known materials with respect to their anharmonicity and perform fully anharmonic {\it ab initio} Green-Kubo calculations for \Ntotal{} of them, finding \Nten\ thermal insulators with $\kappa < 10$\,W/mK including \Nultra\  with ultralow $\kappa \lesssim 1$\,W/mK.
Our analysis reveals that the underlying strong anharmonic dynamics is driven by the exploration of meta-stable intrinsic defect geometries. This is at variance with the frequently applied perturbative approach, in which the dynamics is assumed to evolve around a single stable geometry.
\end{abstract}
\date{\today}
\maketitle
Efficient energy conversion 
is one of the key challenges for the 21$^\text{st}$ century. In this context, the thermal conductivity $\kappa$ is an important material property that can strongly influence device performance and efficiency. This is the case, for example, in combustion engines, where thermal barrier coatings are used to increase operating temperatures~\cite{Perepezko.2009,Clarke.2012}, or in thermoelectric materials for waste-heat recovery~\cite{Snyder.2008}. In both cases, strong thermal insulators are required and the best available materials intrinsically have a low $\kappa < 10$\,W/mK. Additional defect and grain boundary engineering can reduce $\kappa$ even further~\cite{Biswas:2012fw,Ferrando-Villalba.2020}.
Since heat transport is limited by effects beyond the harmonic phonon picture~\cite{Peierls.1929}, elucidating the microscopic anharmonic mechanisms leading to a low intrinsic thermal conductivity is a prerequisite for searching the next generation of thermal insulators. However, our current knowledge about thermal insulators is rather limited. 
Experimental measurements are hindered 
by the challenge of growing macroscopic, well-characterized samples~\cite{Wei.2016}. 
In turn, Springer Materials lists only a little over one thousand measurements for all inorganic solids, and just a few hundred semiconductors and insulators are characterized well at ambient conditions~\cite{Zhu.2021,Gaultois.2013}. 

One way to expand on this knowledge is via computational materials discovery based on (high-throughput) screening of chemical space that is accelerated by artificial-intelligence techniques~\cite{Miller.2017,Chen.2019,Purcell.2022a}.
To this end, reliable simulations of lattice thermal conductivites in the sub-10~W/mK regime are necessary. 
Typically, Peierls-Boltzmann~\cite{Broido.2007,Ravichandran.2018} or Wigner-type~\cite{Simoncelli.2019,Simoncelli.2022} approaches based on a (renormalized~\cite{Aseginolaza.2019}) phonon-picture are employed for this purpose. 
However, these methods are only applicable as long as the phonon quasi-particle picture is valid, i.\,e., when 
the phonon lifetime exceeds the vibration period, as formulated within the Ioffe-Regel limit~\cite{ioffe1960non,Simoncelli.2022}.
In this letter, we demonstrate that the presence of metastable defect geometries is associated with strong anharmonic effects, which break the  Ioffe-Regel limit and result in the exploration of
different energy minima. This breakdown of the phonon picture leads to unexpectedly low thermal conductivities. 

To this end, we employ the \gls{aigk} method which is not affected by the aforementioned limitations since the atomic motion is evaluated via \gls{aimd} simulations~\cite{Marcolongo.2016,Carbogno.2017}. 
Here, we benchmark and validate the approach for \Nexp\ crystalline materials for which reliable experimental measurements are available, showing 
that the necessary accuracy to predict materials in the sub-10~W/mK regime can be reached, even when the phonon picture is no longer valid.
We subsequently devise a hierarchical screening strategy to search for strongly anharmonic thermal insulators in a chemically and structurally diverse space of materials that covers \Nscreening\ experimentally known compounds that are stable at ambient conditions, but without experimentally known $\kappa$.
By that, we identify \Nten\ materials with $\kappa < 10$\,W/mK and we discuss the role of structural displacements to meta-stable geometries 
for the underlying strongly anharmonic dynamics.

\begin{figure}
	\includegraphics[width=\columnwidth]{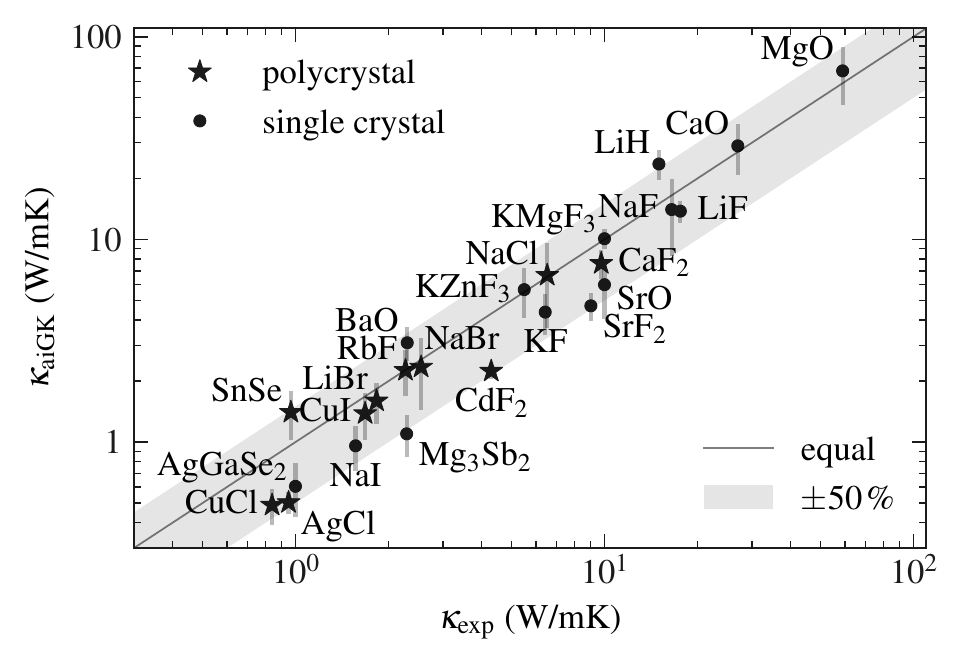}
	\caption{Correlation between measured~($\bullet$ single-crystals, $\bf \star$ polycrystals) and computed~(\gls{aigk}) thermal conductivities at room temperature on a log-log scale. 
        The gray area indicates 50\,\% deviation in terms of $\kappa$ from experiment and the error bars denote the standard error between trajectories. 
        Where multiple experimental references were available, the average was taken.
	}
	\label{fig:kappa.exp}
\end{figure}

The \gls{aigk} approach evaluates the \gls{gk} equation~\cite{Kubo.1957,Kubo.1957oqc}
\begin{align}
  \kappa (T)
    = \frac{V}{3 k_{\rm B} T^{2}} \lim_{t_0 \rightarrow \infty}\int_{0}^{t_0}
    \left\langle {\bf J} (t) \cdot {\bf J} (0) \right\rangle_{T}
    \d t~,
  \label{eq:kappa.GK}
\end{align}
with the heat flux ${\bf J} (t)$, volume $V$, temperature $T$, and Boltzmann constant $k_{\rm B}$. Here, $\langle \bullet \rangle_T$ denotes an ensemble average.
In this work, the heat flux ${\bf J} (t)$ is computed from first principles along \gls{aimd} trajectories using the formalism defined in Ref.~\cite{Carbogno.2017}. 
We use the {\it FHI-aims} code~\cite{Blum.2009,Knuth.2015} and the PBEsol~\cite{Perdew.2008} exchange-correlation functional.   
By neglecting convective contributions to ${\bf J} (t)$, which are usually negligible in solids~\cite{Ladd:1986tv}, physically well-founded strategies to overcome finite-size and -time effects 
can be established.
This includes noise-filtering of the integrand in Eq.\,\eqref{eq:kappa.GK},~i.\,e.,~the \gls{hfacf}, and accounting for long-wavelength excitations missing in the simulation cell via reciprocal-space 
interpolation~\cite{Carbogno.2017} as discussed in detail in a separate work~\cite{Knoop.2022a} and implemented in {\it FHI-vibes}~\cite{Knoop.2020cx}.
Computational parameters were chosen to obtain reliable values of $\kappa$ at room temperature: All \gls{aimd} simulations 
are performed in extended supercells~(160--256 atoms) with trajectory lengths of up to~$60$~ps. Error bars are estimated by 
ensemble averaging over at least three independent trajectories. 
Lattice expansion is accounted for by minimizing the thermal pressure at 300\,K~\cite{Knoop.thesis}. Further details are given in the \gls{sm}~\cite{sm}.

To benchmark the \gls{aigk} approach, we investigate \Nexp\ dielectric compounds, for which 
experimental values of the bulk thermal conductivity $\kappa$ at room temperature are reported in the literature~\cite{sm}.
The reference data covers two orders of magnitude in~$\kappa$, but also carries an uncertainty which is
rooted in differences in sample quality and experimental techniques~\cite{Gaultois.2013}. 
For instance, available data for \gls{snse} varies by up to 50\,\%~\cite{Wei.2016}.
As shown in Fig.~\ref{fig:kappa.exp},
the \gls{aigk} calculations provide a satisfactory agreement with experiment over the whole range of $\kappa$:
For mildly anharmonic conductors with $\kappa > 20$\,W/mK, the quality of the predictions~[\gls{mape} of $22$\,\%] is on par with results from advanced perturbative methods~\cite{Plata.2017,Xia.2020}.
Good accuracy is retained also for strongly insulating materials with $\kappa < 20$\,W/mK, 
which show a comparable \gls{mape} of $28\,\%$ and a resulting \gls{mae} of just 1\,W/mK.
Accounting for strongly anharmonic effects is decisive to reach this accuracy in the sub-10\,W/mK regime, as exemplified by the fact that 
the \gls{aigk} method 
correctly predicts $\kappa$ to be 1.4\,W/mK for \gls{cui}, just 16\,\% smaller than in experiment. In contrast, different phonon-based approaches yield values between 5.8 and 6.8 W/mK, overestimating experiment 
by $\gtrsim$ 350\%, cf. \gls{sm} Sec.\,IV. Even more importantly, the accurate lifetimes obtained from fully anharmonic \gls{aimd} reveal a breakdown of the phonon picture in this compound, cf. \gls{sm} Sec.\,VI.
\begin{figure}
	\includegraphics[width=\columnwidth]{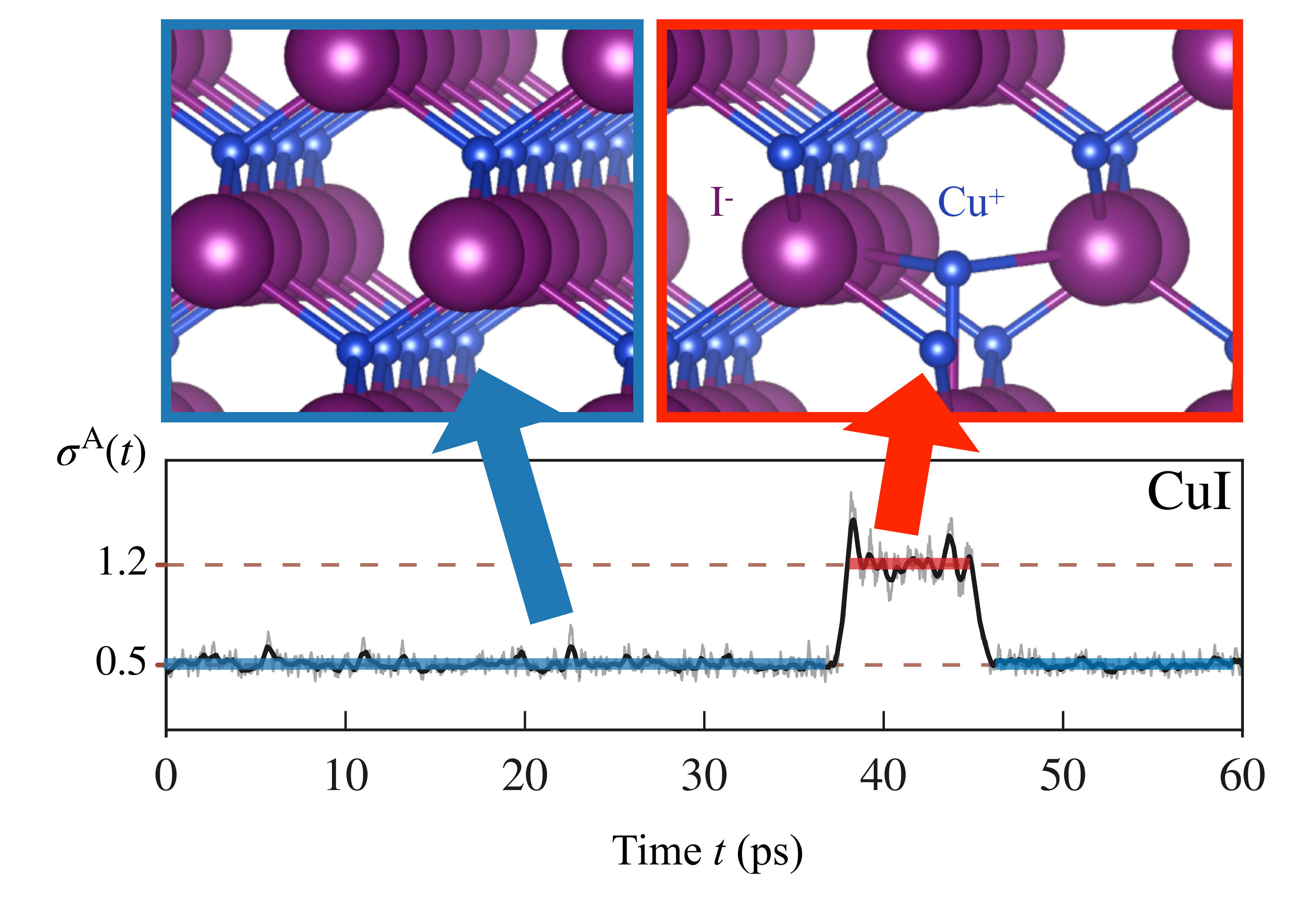}
	\caption{Anharmonicity measure $\sigmaAt$ for \gls{cui} evaluated at each time step during one \gls{aimd} trajectory. The upper row shows averaged positions in $(10\bar{1})$ direction for time-periods with $\sigmaAt \approx 0.5$ (blue), and increased $\sigmaAt \approx 1.2$ (red), in which one Cu cation moves to an interstitial site along $(\bar{1}11)$.
	}
	\label{fig:CuI}
\end{figure}
\begin{figure*}
  \includegraphics[width=\textwidth]{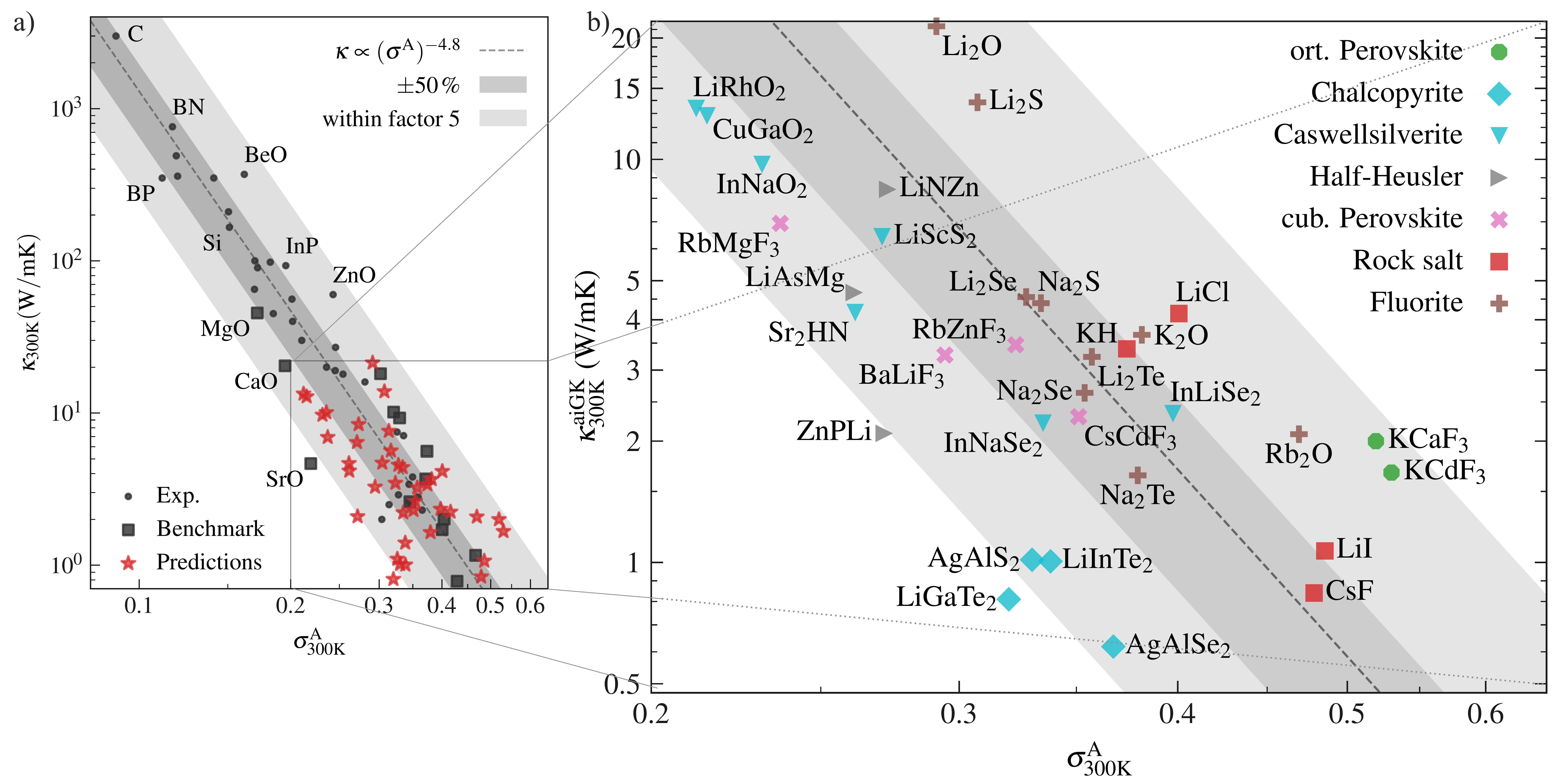}
	\caption{Thermal conductivity $\kappa$ vs. anharmonicity at 300\,K, $\sigmaAtot$ (tabulated data given in the \gls{sm}~\cite{sm}). The dashed trend line and grey areas are shown to guide the eye.
        a) The qualitative scaling of $\kappa$ with $\sigmaAtot$ suggested in Ref.~\cite{Knoop.2020} and used to disregard good thermal conductors during screening
        is showcased using experimental measurements~(black $\bullet$) as well as our \gls{aigk} calculations~[black $\blacksquare$: benchmarks from Fig.\,\ref{fig:kappa.exp}; red {\color{red} $\bf \star$}: predictions from subplot~(b)]. 
	b) Thermal conductivity \gls{aigk} predictions for materials with $\sigmaAtot > 0.2$. The symbol and color code indicate the lattice type.
	}
	\label{fig:kappa}
\end{figure*}

For \gls{cui}, an analysis of the \gls{aimd} dynamics confirms that strong anharmonic effects are at play:
Figure\,\ref{fig:CuI} shows the time evolution of the anharmonicity measure~$\sigmaAt$ introduced in Ref.~\cite{Knoop.2020}, 
which quantifies the differences between the actual and the harmonic \gls{pes} and 
hence would vanish in perfectly harmonic materials~\cite{Knoop.2020,sm}.
The sudden increase of $\sigmaA (t)$ from its value of 0.5 
to 1.2 signals that the harmonic approximation 
becomes qualitatively incapable to describe the dynamics:
In this several picoseconds long period, a meta-stable Frenkel defect is formed~\cite{Frenkel.1926}.
As shown in  Fig.\,\ref{fig:CuI}, one Cu~cation moves to an interstitial site along the $(\bar{1}11)$~direction 
and vibrates around this meta-stable defect geometry, before it jumps back to the zincblende equilibrium configuration 
associated with~$\sigmaAt \approx 0.5$. 
This temporary defect formation may be viewed as a local precursor of the superionic $\beta$-phase of \gls{cui},
which is known to exhibit a considerable defect concentration $>10\,\%$,
but only becomes stable at significantly higher temperatures~($>$643\,K)~\cite{Keen.1995}. 

In the next step, we employ the \gls{aigk} method to screen for thermal insulators via a hierarchical workflow: We study \Nscreening\ experimentally known binary and ternary materials for which $\kappa$ at room temperature has not been measured. This set of materials covers seven space groups~(62, 122, 166, 186, 216, 221, 225), including not only simple rock~salt materials with two atoms per unit cell, but also more complex compounds,~e.\,g.,~orthorombic perovskites with 20~atoms in the unit cell. Note that only materials with elements lighter than Lanthanum are considered to avoid potential artifacts in the dynamics associated to the treatment of spin-orbit coupling. First, we compute the band gap and the anharmonicity measure at 300\,K, $\sigmaAtot$, for all \Nscreening\ compounds. The results cover the range $\sigmaAtot = 0.1-0.7$, cf.~\gls{sm}. Over 50\% of the materials feature an anharmonicity of $\sigmaAtot \approx 0.23$ or less, in line with Ref.~\cite{Knoop.2020}. Second, we exclude materials with DFT-PBEsol band gaps below 0.2\,eV to ensure that electronic contributions to heat transport are negligible and we discard more harmonic materials with $\sigmaAtot < 0.2$ to avoid unnecessary and expensive calculations, since those materials can be expected to be good thermal conductors~\cite{Knoop.2020} that can be treated with perturbative approaches. The first aspect is substantiated by Fig.\,\ref{fig:kappa}a, which shows that the materials with $\sigmaAtot < 0.2$ feature $\kappa$ values (much) larger than 10~W/mK. 
From the remaining $\sigmaAtot > 0.2$~pool, we generate a diverse test set of 
materials that uniformly cover the range $\sigmaAtot = 0.2-0.6$. Materials with a particularly slow phonon dynamics were excluded to ensure that simulation time convergence can be reached within reasonable trajectory lengths~$\le 60$\,ps, as detailed in the SM~\cite{sm}. Finally, we obtain thermal conductivities for \Nnew{}~materials for which $\kappa$ has \emph{not} been measured yet, see~Fig.\,\ref{fig:kappa}b.

\begin{figure}
	\includegraphics[width=\columnwidth]{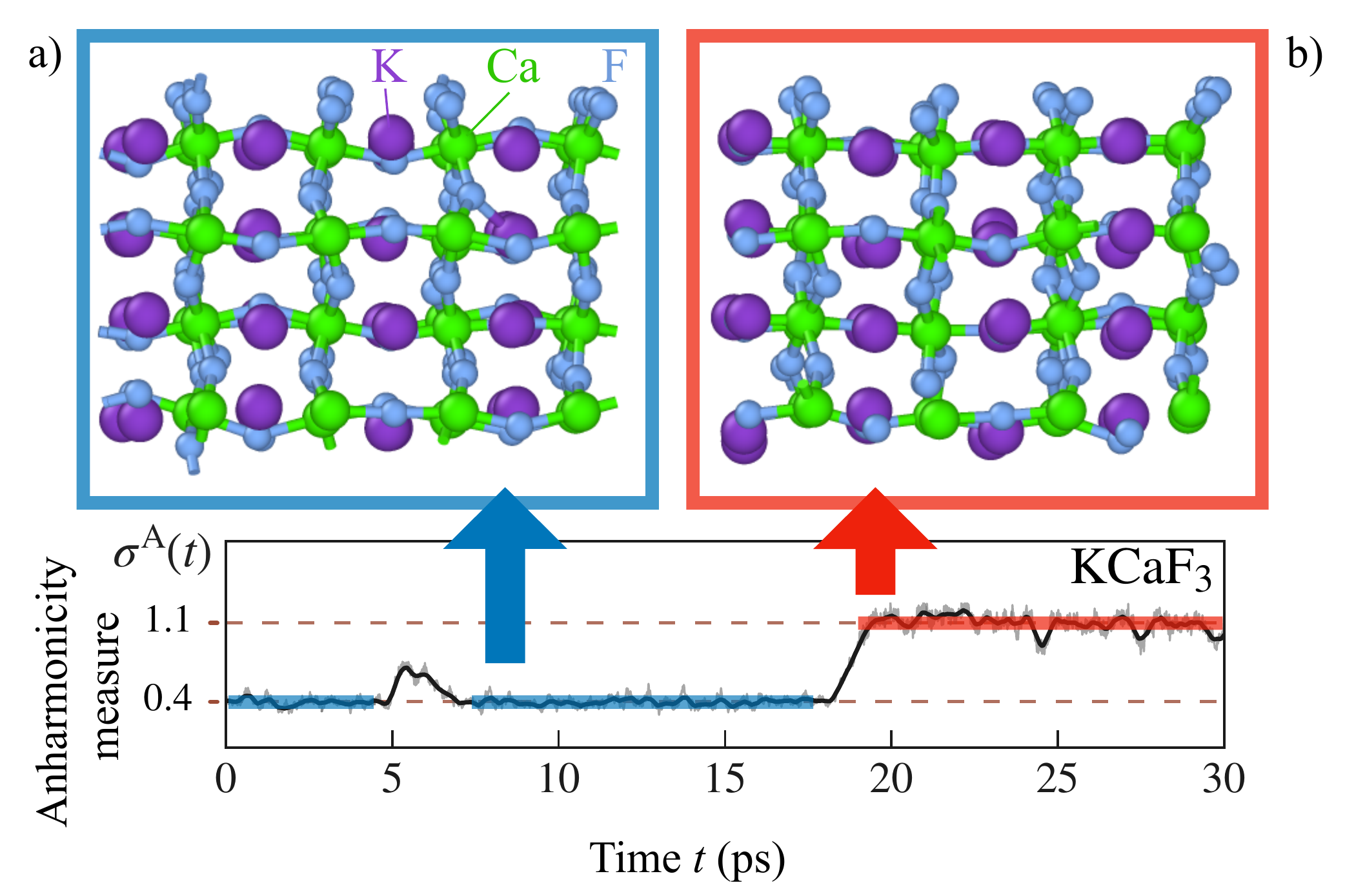}
	\caption{Anharmonicity measure $\sigmaAt$ for \kcaf{} evaluated at each time step during one \gls{aimd} trajectory. The upper row shows time-averaged positions in (100) direction for the periods with a)~$\sigmaA (t) \approx 0.4$ (blue), and b)~$\sigmaA (t) \approx 1.1$ (red). Note the rearrangement of Ca-F bonds from zigzag to straight shape when comparing a) and b). See \gls{sm} for a video of the tilt rearrangement around 18\,ps.}
	\label{fig:KCaF3}
\end{figure}

Although the $\sigmaA$-criterion is only qualitatively valid for the identifcation of thermal insulators,  no good thermal conductors are found within this set.
On the contrary, we find \Nten\ new materials with a thermal conductivity of $\kaigk < 10\,{\rm W/mK}$, \None\ of which with an ultralow conductivity $\lesssim 1\,{\rm W/mK}$.
These compounds comprise simple binary, cubic materials such as the rock salts CsF and LiI or the fluorite Na$_2$Te, but also more complex structures such as the KCdF$_3$ and \kcaf{} perovskites.
Furthermore, we find four chalcopyrites~(\agalse{}, \agals{}, \ligate{}, \liinte{}), a material class that has recently gained interest for thermoelectric applications~\cite{Zhu.2019,Plata.2022,Xie.2022}.
In the strongly insulating regime shown in Fig.\,\ref{fig:kappa}b, the simple power-law correlation between $\kappa$ and $\sigmaAtot$ observed in 
Fig.\,\ref{fig:kappa}a is no longer valid due to the increased variety and complexity of the compounds.
This can be explained by comparing two materials~(\ligate, \kcaf{}); more details are given in the \gls{sm}~\cite{sm}: 
The strongly insulating character of the chalcopyrite~\ligate\ ($\kaigk = 0.8$\,W/mK) can already be rationalized in a 
harmonic picture, 
since most phonon branches of this compound, including the acoustic ones, exhibit only an extremely weak dispersion, cf.~Suppl.\,Fig.\,1. 
Accordingly, only few modes
with $\omega < 4$\,THz can contribute to 
thermal transport, and even these feature very low group velocity $\lesssim 1.5\,$km/s, cf.~Suppl.\,Fig.\,2. Since this is a common property in chalcopyrites, Plata {\it et al.} were able to correctly predict thermal insulators in
this material class~\cite{Plata.2022}, even if the employed perturbative approach 
might overestimate $\kappa$ due to the low-order description of anharmonicity.
In comparison, \kcaf{} features much more dispersive phonon frequencies: The majority of modes across the whole frequency range
contributes to heat transport with considerably larger group velocities that reach $5\,$km/s, suggesting a moderately good thermal conduction. In this case, the
strongly insulating character of \kcaf{} ($\kaigk = 2$\,W/mK) can only be explained by its strong anharmonicity~($\sigmaAtot = 0.5$), which
results in extremely low phonon lifetimes $\lesssim 1$\,ps.
As shown in Fig.\,\ref{fig:KCaF3}, this is reflected in the real-space dynamics that displays temporary rearrangements 
towards different, meta-stable configurations. These involve an increase of $\sigmaAt$, as discussed above for \gls{cui} in Fig.~\ref{fig:CuI}.
Unlike \gls{cui} however, in which these rearrangements correspond
to the formation of intrinsic point defects, 
we observe more extended distortions in \kcaf{} that span 
multiple primitive unit cells, cf.~Fig.~\,\ref{fig:KCaF3}.
As discussed in the \gls{sm}, these configurations are associated to reorientations of the CaF$_3$ octahedra into different, nearly-degenerate octahedral tilting patterns.
These meta-stable configurations are a common feature of perovskite materials and their dynamics often leads to the stabilization of higher-temperature phases~\cite{Bertolotti.2017,Klarbring.2019}. For \kcaf{}, the described dynamics is a precursor of the cubic phase, which becomes stable at temperatures $> 550$\,K~\cite{Bulou.2011}.

Note that these temporary atomic rearrangements into meta-stable geometries are not observed in all trajectories. Nonetheless, their microscopic origin,~i.\,e.,~a strongly-anharmonic \gls{pes} featuring 
multiple minima, influences the whole dynamics. Even when these minima are not explicitly explored, harmonic and anharmonic contributions are at least of equal importance, 
as signaled by the high $\sigmaA$ value $\geq 0.5$, and the fact that the fully anharmonic lifetimes, which violate the Ioffe-Regel limit in both compounds, demonstrate a breakdown of the phonon picture, 
as shown in Suppl.\,Fig.\,3 and Sec.~VI in the \gls{sm}.
These observations are direct evidence that the existence of barely accessible meta-stable states is suﬃcient to increase the anharmonicity in \gls{cui} and \kcaf{}  beyond the level at which it can be treated as a perturbation.
We also note that these localized effects are different from other local, anharmonic scattering mechanisms such as ``rattlers'' that do not involve temporal lattice distortions~\cite{Tadano.2018n2c}.

In summary, we presented a comprehensive set of aiGK calculations to search for thermal insulators.
By hierarchically screening over \Nscreening\ experimentally known materials,
we identified \Nten\ new materials with $\kappa < 10$\,W/mK including \Nultra\ compounds with ultralow $\kappa \lesssim 1$\,W/mK. 
Our analysis reveals that the dynamics in many thermal insulators is not limited to displacements around a \emph{single}, stable equilibrium geometry, as assumed in perturbative approaches like the \gls{bte}. On the contrary, strongly anharmonic dynamical effects such as point defect formation in \gls{cui} or extended rearrangements in \kcaf{} involve multiple meta-stable geometries that need to be accounted for. Besides providing a microscopic explanation
of the insulating character of these materials, our findings suggest a fundamentally different route for reducing $\kappa$ that is complementary to traditional phonon-based approaches~\cite{Garg.2011,McGaughey.2006}: For instance, incorporating defects that promote the formation of additional meta-stable states~\cite{Carbogno:2014wa,Dobrovolsky.2017} can strengthen the anharmonic effects leading to low thermal conductivity even at temperatures at which these states are not fully accessible.
Our results and the more than 10~nanoseconds of available PBEsol--\gls{aimd} trajectories can be used to benchmark the various methods available for computing $\kappa$ in increasingly anharmonic materials. Similarly, this data lends itself to 
train machine-learning potentials to accelerate $\kappa$ predictions in the future~\cite{Sosso.2012,Korotaev.2019,Verdi.2021} and to develop improved descriptors for $\kappa$~\cite{Miller.2017,Chen.2019,Purcell.2022a}. In this regard, quantitative
models that go beyond the qualitative $\sigmaA$ scaling used in this work to identify anharmonic materials can accelerate materials' space exploration considerably. 

The electronic-structure theory calculations and \gls{aimd} trajectories produced in this project are available via the NOMAD repository~\cite{nomad1}. 
Resources for the plots can be found on figshare~\cite{figshare}. 
\begin{acknowledgments}
This project was supported by the NOMAD Center of Excellence (European Union’s Horizon 2020 research and innovation program, grant agreement No. 951786), the ERC Advanced Grant TEC1p (European Research Council, grant agreement No. 740233), the North-German Supercomputing Alliance (HLRN), BigMax (the Max Planck Society’s Research Network on Big-Data-Driven Materials-Science). We acknowledge PRACE for awarding us access to JUWELS at GCS@FZJ, Germany. F.K. acknowledges support from the Swedish Research Council (VR) program 2020-04630, and the Swedish e-Science Research Centre (SeRC). T.P. would like to thank the Alexander von Humboldt Foundation for their support through the Alexander von Humboldt Postdoctoral Fellowship Program.
\end{acknowledgments}

\bibliography{literature}
\bibliographystyle{apsrev4-1}

\end{document}